\documentstyle[12pt,titlepage]{article}
\input epsfig.sty

\font\grande=cmr9.5 scaled \magstep4
\font\medio=cmr9.5 scaled \magstep2
\outer\def\beginsection#1\par{\medbreak\bigskip
      \message{#1}\leftline{\bf#1}\nobreak\medskip
\vskip-\parskip
      \noindent}

\def\dis{\displaystyle}
\newcommand{\mpl}{M_{\rm Pl}}
\newcommand{\beq}{\begin{equation}}
\newcommand{\eeq}{\end{equation}}

\newcommand{\beqn}{\begin{eqnarray}}
\newcommand{\eeqn}{\end{eqnarray}}
\newcommand{\nbeqn}{\begin{eqnarray*}}
\newcommand{\neeqn}{\end{eqnarray*}}
\newcommand{\bcen}{\begin{center}}
\newcommand{\ecen}{\end{center}}
\def\laq{\raise 0.4ex\hbox{$<$}\kern -0.8em\lower 0.62
ex\hbox{$\sim$}}
\newcommand{\benu}{\begin{enumerate}}
\newcommand{\eenu}{\end{enumerate}}
\newcommand{\bite}{\begin{itemize}}
\newcommand{\eite}{\end{itemize}}
\newcommand{\bdes}{\begin{description}}
\newcommand{\edes}{\end{description}}
\newcommand{\bdis}{\begin{displaymath}}
\newcommand{\edis}{\end{displaymath}}
\newcommand{\bary}{\begin{array}}
\newcommand{\eary}{\end{array}}

\def\gaq{\raise 0.4ex\hbox{$>$}\kern -0.7em\lower 0.62
ex\hbox{$\sim$}}

\begin{document}
\bibliographystyle {unsrt}

\titlepage

\vspace{15mm}
\begin{center}
{\grande Response of VIRGO detectors to pre-big-bang gravitons }\\
\vspace{15mm}
D. Babusci $^a$ and M. Giovannini $^b$ 
\vspace{15mm}

{\sl $^a$ INFN- Laboratori Nazionali di Frascati, 1-00044 Frascati, 
Italy}\\

{\sl $^b$ Institute for Theoretical Physics, Lausanne University, }\\
{\sl BSP-Dorigny, CH-1015, Switzerland}
\end{center}

\vskip 2cm
\centerline{\medio  Abstract}

\noindent
{The sensitivity achievable by a pair of VIRGO detectors to
stochastic and isotropic gravitational wave
backgrounds produced in pre-big-bang models is discussed
in view of the development of a second VIRGO interferometer.
We describe a semi-analytical
technique allowing to compute the signal-to-noise ratio for
(monotonic or non-monotonic)
logarithmic energy spectra of relic gravitons of arbitrary
 slope. We apply our results to the case of two correlated and
coaligned VIRGO detectors  and we compute their achievable
sensitivities.
 We perform our calculations both for
the usual case of minimal string cosmological scenario and
in the case of
a non-minimal scenario (originally suggested by Gasperini)
 where a long dilaton dominated phase is
present prior to the onset of the ordinary
radiation dominated phase. In this framework, we investigate possible
improvements of the achievable sensitivities by selective
reduction of the thermal contributions
(pendulum and pendulum's internal modes)
to the noise power spectra of the detectors.
Since a reduction of the shot noise does not increase significantly
the expected sensitivity of a VIRGO pair (in spite of the
relative spatial location of the two detectors)
our findings support the experimental efforts directed towards a substantial
reduction of thermal noise.
\vspace{5mm}

\newpage

\renewcommand{\theequation}{1.\arabic{equation}}
\setcounter{equation}{0}

\section{The problem and its motivations}

Every variation of the background
geometry produces graviton pairs which are stochastically
distributed and whose logarithmic energy spectra represent
a faithful snapshot of the (time) evolution of the curvature
scale at very early times \cite{1}. Indeed, one of the
peculiar features of stochastic graviton backgrounds is that
their energy spectra extend over a huge interval of (present)
frequencies.  Since gravitational interactions
are much weaker than electromagnetic interactions they also
decouple much earlier. Therefore the logarithmic energy
spectra of relic gravitons produced by the pumping action of
the gravitational field can very well extend for (approximately)
twenty five orders of magnitude in frequency \cite{3}. From the
physical point of view, this observation implies that the energy
spectra of relic gravitons can be extremely relevant in order to
probe the past history of the Universe in a regime which will
never be directly accessible with observations of electromagnetic
backgrounds of cosmological origin \cite{2}.

In spite of the fact that the {\em nature}
of the production mechanism is shared by different
types of models \cite{1}, the specific {\em amplitudes of the
energy spectra} can very well change depending
upon the behavior of the background evolution.
An example in this direction are
logarithmic energy spectra increasing in frequency \cite{9} or even
non-monotonic spectra originally discussed by Gasperini \cite{gas1}.

Different theoretical signals (with different
spectral distributions) lead to
 detector outputs of different amplitudes.
Therefore, in order to evaluate the performances
of a given detector one has to choose the specific
functional form of the logarithmic energy spectrum.
String cosmological models \cite{10} are an interesting
theoretical laboratory leading usually to sizable theoretical
signals in the operating window of wide band interferometers (WBI)
\cite{11}. A possible
detection of these backgrounds would represent an
interesting test for cosmological models
inspired by the low energy string effective action.
Other possible choices are represented
by scale invariant spectra \cite{gri,8,gri2} or
by tilted (``blue'' \cite{gri2}) spectra whose energetical content
is typically concentrated at frequencies larger than
the mHz \cite{gio2,noi}.

The signal induced in the detector output by a stochastic
background of gravitational radiation
 is indistinguishable from the intrinsic noise of the
detector itself. This implies that, unless the amplitude of the signal
is very large,
the only chance of direct detection of these backgrounds
lies in the analysis of the
correlated fluctuations of the outputs of, at least, two detectors
affected by independent noises\footnote{We stress that
(see, for instance the second paper in Ref. \cite{chr}) the ratio
between the minimum signal detectable in the cross-correlation and the
minimum signal detectable in the case of a single detector is
$\sim 1/\sqrt{T\,\Delta f}$ where $T$ represents the duration of
the correlation experiment and $\Delta f$ is the width of the frequency
band probed by the detectors. This means that for a measurement with
$\Delta f = 1 $ kHz (like in the case of the VIRGO interferometer) and
$T = 1$ yr , the minimum signal detectable with a correlation is $10^{5}$
times smaller than in the case of a single detector.}.

The problem of the optimal processing of the detector outputs required
for the detection of the stochastic background has been considered by
various authors \cite{mic,chr} and it was also reviewed in Ref. \cite{alr}.
Suppose, indeed, that the  signal registered at each detector can be
written as (we limit ourselves to the case of two detectors
$(i\,=\,1,2)$)
\begin{equation}
s_{i}\,=\,h_{i}(t)\,+\,n_{i}(t)\,,
\end{equation}
where we have indicated with $n$ the intrinsic noise of the detector,
and with $h$ the gravitational strain due to the stochastic background.
By assuming that the detector noises are stationary and uncorrelated,
the ensemble average of their Fourier components satisfies
\begin{equation}
\langle n^{\ast}_{i}(f)\,n_{j}(f')\rangle\,=\,\frac{1}{2}\,\delta(f-f')
\,\delta_{ij}\,S^{(i)}_{n}(|f|)\;,
\end{equation}
where $S_{n}(|f|)$ is usually known as the one-sided noise power spectrum
and is expressed in seconds. Starting to the signals $s_1$ and $s_2$, a
correlation ``signal'' for an observation time $T$ can be defined in the
following way:
\begin{equation}
S\,=\,\int_{- T/2}^{T/2}\,{\rm d} t\,\int_{- T/2}^{T/2}\,{\rm d} t'
\,s_1 (t)\,s_2 (t')\,Q (t - t')
\end{equation}
where $Q$ is a filter function that depends only by $t - t'$ because we
assume that $n$ and $h$ are both stationary. The optimal choice of $Q$
corresponds to the maximization of the signal-to-noise ratio associated
to the ``signal'' $S$.
 In this calculation the stochastic background,
besides to be stationary, is also assumed to be isotropic, unpolarized
and Gaussian.
Under the further assumptions that detector noises
are Gaussian, much larger in amplitude than the gravitational strain and
statistically independent on the strain itself, it can be shown
\cite{mic,chr,alr} that the signal-to-noise ratio in a frequency range
$(f_{\rm m},f_{\rm M})$ is given by\footnote{ For a clear discussion
about these assumptions see Ref. \cite{alr}. In this
paper are also discussed the modifications needed in the case in which
most of these assumptions are relaxed. Finally, in order to avoid
possible confusions we stress that the definition of
the SNR is the one discussed in \cite{noi} and it is essentially the square
root of the one discussed in \cite{mic,chr,alr}.}:
\begin{equation}
{\rm SNR}^2 \,=\,\frac{3 H_0^2}{2 \sqrt{2}\,\pi^2}\,F\,\sqrt{T}\,
\left\{\,\int_{f_{\rm m}}^{f_{\rm M}}\,{\rm d} f\,
\frac{\gamma^2 (f)\,\Omega_{{\rm GW}}^2 (f)}{f^6\,S_n^{\,(1)} (f)\,
S_n^{\,(2)} (f)}\,\right\}^{1/2}\; ,
\label{SNR}
\end{equation}
where $H_0$ is the present value of the Hubble parameter and $F$ is a
numerical factor depending upon the geometry of the two detectors.
In the case of the correlation between two interferometers $F= 2/5$.
In Eq. (\ref{SNR}),
the  performances achievable by the pair of detectors are certainly
controlled by the noise power spectra (NPS) $S_n^{\,(1,2)}$.
However in Eq. (\ref{SNR}),
on top of NPS, there are two important quantities.
The first one is the {\em theoretical}
background signal defined through the logarithmic energy spectrum
(normalized to the critical density $\rho_c$) and expressed at the present
(conformal) time $\eta_0$ \footnote{In most of our equations we drop the
dependence of spectral quantities upon the present time since all the
quantities introduced in this paper are evaluated today.}
\begin{equation}
\Omega_{{\rm GW}}(f,\eta_0)\,=\,\frac{1}{\rho_{c}}\,
\frac{{\rm d} \rho_{{\rm GW}}}{{\rm d} \ln{f}}\,=\,
\overline{\Omega}(\eta_0)\,\omega(f,\eta_0)\,.
\label{1}
\end{equation}
As we can see, we have chosen to parametrize the theoretical spectrum
through a time-independent amplitude ($\bar{\Omega}(\eta_0)$) and a
frequency-dependent part ($\omega(f)$).
The important quantity appearing in eq. (\ref{SNR})
is the overlap reduction function $\gamma(f)$ \cite{chr,alr}
which is a dimensionless function describing the reduction in  the sensitivity
of the two detectors (at a given frequency $f$)
arising from the fact that the two detectors are not in the
same place and, in general, not coaligned
(for the same location and orientation $\gamma (f) = 1$). Since
the overlap reduction function cuts-off
the integrand of Eq. (\ref{SNR}) at a frequency which approximately
corresponds to the inverse separation between the two detectors, it may
represent a problematic (but controllable) element in the reduction
of the sensitivity of a given pair of detectors.

Various ground-based interferometric detectors are presently under
construction (GEO \cite{geo}, LIGO-LA, LIGO-WA \cite{lig}, TAMA \cite{tam},
VIRGO \cite{vir}). Among them, the pair consisting of most homogeneous
(from the point of view of the noise performances) detectors with minimum
separation is given by the two LIGOs (VIRGO and GEO are even closer, but
they have different performances for what concerns the NPS). However, this
separation ($\simeq$ 3000 km) is still too large. The overlap
reduction function $\gamma (f)$ for the pair
LIGO-LA$-$LIGO-WA encounters
its first zero at 64 Hz,  falling off (swiftly) at higher
frequencies, i.e., right in the region where the two LIGOs, at least in
their initial version, have better noise performances.

The motivation of our exercise is very simple. We want to study
the sensitivity of a system of two VIRGO-like detectors to stochastic
backgrounds of gravitational radiation.
Up to now, if we exclude the case of Ref. \cite{noi},
the correlation between two VIRGO detectors has not been seriously explored
in the literature.
The reason for this lack of studies is that, in contrast with the
LIGO project where two detectors are simultaneously under construction,
 only one  VIRGO detector is currently being built. However, recently,
within the European gravitational
wave community, the possibility of building in Europe an interferometric
detector of dimensions comparable to VIRGO has  received
close attention \cite{gia}. Therefore, there is a chance
that in the near future the VIRGO detector,
now under construction at Cascina (Pisa) in Italy, will be complemented
by another interferometer of even better performances very close
(at a distance $d\,<\,1000$ km) to it. In
this paper we examine in detail the possible improvements in the VIRGO
sensitivity as a result of direct correlation of two VIRGO-like detectors.
Technological improvements in the construction of the
interferometers can be reasonably expected in the next years. Thus
the VIRGO detectors will gradually
evolve towards an advanced configuration \cite{gia}.
For this reason we also examine
the possible consequences of a selective improvements of the noise
characteristics of the two detectors on the obtained results.
In order to make our analysis
concrete we will pay particular attention to the evaluation
of the performances of a pair of VIRGO detectors in the case
of string cosmological models \cite{10,11}.

In order to evaluate precisely the performances of
a pair of VIRGO detectors we will use the following
logic. First of all we will pick up a given class
of theoretical models which look particularly
promising in view of their spectral properties
in the operating window of the WBI.
Secondly we will analyze the signal-to-noise
ratios for different regions of the parameter
space of the model. Finally we will
implement some selective reduction of the noises
and we will compare the results with the ones
obtained in the cases
where the noises are not reduced. We will
repeat the same procedure for different classes
of models.

The results and the investigations we are
reporting can be applied to spectra of arbitrary
functional form. The only two requirements
we assume will be the continuity of the
logarithmic energy spectra (as a function of the present
frequency) and of their first derivative. We will also give
some other examples in this direction.

The plan of our paper is then the following. In Section II
we introduce the basic semi-analytical tecnique which allows
the evaluation of the SNR for a pair of WBI. In Section III
we will evaluate the performances of a pair of VIRGO detectors
in the case of string cosmological models.
In Section IV we will show how to implement
a selective noise reduction and we will investigate the
impact of such a reduction in the case
of the parameter space of the models previously analyzed.
Section V contains our final discussion and the
basic summary of our results.

\renewcommand{\theequation}{2.\arabic{equation}}
\setcounter{equation}{0}
\section{SNR evaluation}

In the operating window of the VIRGO detectors the theoretical
signal will be defined through the logarithmic energy spectrum
reported in Eq. (\ref{1}). In the present Section we shall not make any
specific assumption concerning $\omega(f)$ and our results have
general applicability. We will only assume that it is a continuous
function of the frequency and we will also assume that its first
derivative is well defined in the operating window of WBI.
This means that $\omega(f)$ can be, in principle, a non-monotonic
function.

\subsection{Basic Formalism}

The expected noise power spectrum for the VIRGO detector \cite{vir}
is well approximated by the analytical fit of Ref. \cite{cuo}, namely
\begin{equation}
\Sigma_n (f)\,=\,\frac{S_n (f)}{S_0}\,=\,
\left\{
\begin{array}{lc}
\infty &\qquad f < f_b \\ [8pt]
\dis \Sigma_1\,\biggl(\frac{f_{{\rm a }}}{f}\biggr)^5\,+\,
\dis \Sigma_2\,\biggl(\frac{f_{{\rm a}}}{f}\biggr)\,+\,
\dis \Sigma_3\,\biggl[ 1 + \biggl(\frac{f}{f_{\rm a}}\biggr)^2\biggr],&
\qquad  f \ge f_b
\label{NPS}
\end{array}
\right.
\end{equation}
with
\bdis
S_0\,=\,10^{-44}\,{\rm s}\;,\qquad f_a\,=\,500\,{\rm Hz}\;,
\qquad f_b\,=\,2\,{\rm Hz}\;,\qquad
\begin{array}{l}
\Sigma_1\,=\,3.46\,\times\,10^{-6} \\
\Sigma_2\,=\,6.60\,\times\,10^{-2} \\
\Sigma_3\,=\,3.24\,\times\,10^{-2}\,.
\end{array}
\edis
In order to compute reliably (and beyond naive power counting
arguments) the SNR we have to specify the overlap reduction
function $\gamma(f)$. The relative location and orientation of the
two detectors determines the functional form of $\gamma(f)$
which  has to be gauged in such a way that the
overlap between the two detectors
is maximized (i.e. $\gamma(f) \simeq 1$ for most of the
operating window of the two VIRGO). Moreover, the two interferometers
of the pair
should also be sufficiently far apart in order to decorrelate the
local seismic and electromagnetic noises. Since the precise
location of the second VIRGO detector has not been specified so
far \cite{gia}, we find useful to elaborate about this point by computing
the overlap reduction functions corresponding to two coaligned
VIRGO interferometers with different spatial separations. The results of 
these calculations are reported in Fig. \ref{over}\footnote{ For 
illustrative purposes, we assumed that a distance of about 50 km is 
sufficient to decorrelate local seismic and e.m. noises. Such a 
hypothesis is certainly justified within the spirit of our exercise.}.
\begin{figure}[!ht]
\centerline{\epsfxsize = 6.2 cm  \epsffile{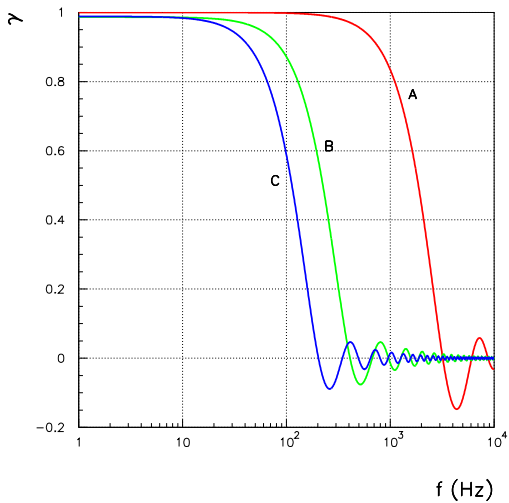}}
\vspace*{-0.5cm}
\caption[a]{We report the overlap reduction function(s) for the correlation of
the VIRGO detector presently under construction in Cascina
(43.6 N, 10.5 E) with a coaligned interferometer whose (corner)
station is located at: A) (43.2 N, 10.9 E), $d\,=\,58$ km (Italy);
B) (43.6 N, 4.5 E), $d\,=\,482.7$ km (France); C) (52.3 N, 9.8 E),
$d\,=\,958.2$ km (Germany). The third site (C) corresponds
to the present location of  the GEO detector. Notice that from A to C
the position of the first zero of $\gamma(f)$ gets shifted in the 
infra-red.}
\label{over}
\end{figure}
Needless to say that these choices are purely theoretical and are
only meant to illustrate the effects of the distance
on the performances of the VIRGO pair.

The curves labeled with A, B, and C shown in Fig. \ref{over}
correspond to different distances $d$ between the site
of the VIRGO detector (presently under construction
in Cascina, near Pisa) and the central corner station
of a second coaligned VIRGO interferometer. Let us now look at
the position of the first frequency $f_i$ for which
$\gamma(f_i)\,=\,0$ for each of the curves. We can notice
that by increasing  $d$ (i.e., going from A to C)
the value of $f_i$ gets progressively shifted towards lower and lower
frequencies,  linearly with $d$. This
means that, for the specific purpose of the detection
of a stochastic background of gravitational radiation, the position
of the first zero of the overlap reduction function cannot
be ignored. In general we would like $f_i$ to be
slightly larger than the frequency region where the sensitivity
of the pair of wide band detectors is maximal.  In the explicit examples
presented in this paper we will focus our attention
on the case A. The other two configurations have been the subject of a
related investigation \cite{noi2}.

\subsection{SNR and bounds on the graviton spectrum}

By inserting the parametrization (\ref{1}) into  Eq. (\ref{SNR})
we can write
\begin{equation}
{\rm SNR}^2 \,=\,\frac{3 H_0^2}{5 \sqrt{2}\,\pi^2}\;
\sqrt{T}\;\frac{\overline{\Omega}}{f_0^{5/2}\,S_0}\;J \;,
\label{snrrescaled}
\end{equation}
where we introduced the (dimension-less) integral
\begin{equation}
J^2 \,=\,\int_{\nu_{\rm m}}^{\nu_{\rm M}}\,{\rm d} \nu\,
\frac{\gamma^2\,(f_0 \nu)\,\omega^2(f_0 \nu)}
{\nu^6\,\Sigma_n^{\,(1)} (f_0 \nu)\,
\Sigma_n^{\,(2)} (f_0 \nu)}\;.
\label{Jint}
\end{equation}
Here the integration variable is $\nu\,=\,f/f_0$, with $f_0$ a
generic frequency scale within the operating window of the
interferometer, and the integration domain is restricted to
the region $f_{\rm m}\,\le\,f\,\le\,f_{\rm M}$
(i.e., $\nu_{\rm m}\,\le\,\nu\,\le\,\nu_{\rm M}$).
In the following we will choose $f_0\,=\,100$ Hz and, taking into
account the frequency behavior of $\gamma (f)$ (see Fig. \ref{over}),
we can assume  $f_{\rm M}\,=\,10$ kHz (i.e., $\nu_{\rm M}\,=\,100$).
The lower extreme $f_{\rm m}$ is put equal to the frequency $f_b$
entering Eq. (\ref{NPS}) (i.e., $\nu_{\rm m}\,=\,0.02$).

For the chosen values of $f_0$ and $S_0$ (see Eq. (\ref{NPS}))
one has  ($H_0\,=\,100\,\times\,h_0\;{\rm km}\,\cdot\,
{\rm s}^{-1}\,\cdot\,{\rm Mpc}^{-1}$):
\begin{equation}
h_{0}^2\,\overline{\Omega}\,\simeq\,\frac{4.0\,\times\,10^{-7}}{J}\;
\left(\,\frac{1\;{\rm yr}}{T}\,\right)^{1/2}\;{\rm SNR}^2\;.
\label{sens}
\end{equation}
Since we will often refer to this formula we want to
stress its physical meaning. Suppose that the functional form of $\omega (f)$
is given. Then the numerical value of the
integral $J$ can be precisely computed and, through Eq. (\ref{sens}),
$\overline{\Omega}$ can be estimated.
This quantity, inserted in Eq. (\ref{1}),
determines for each frequency $f$ the minimum $\Omega_{\rm GW}$
detectable (for an observation time $T$, with a signal-to-noise
ratio SNR) by the correlation of the two detector outputs.

In the next section, $\overline{\Omega}$ will be compared with
two other quantities: $\overline{\Omega}^{\,{\rm th}}$ and
$\overline{\Omega}^{\,{\rm max}}$. The first is the theoretical
value of the normalization of the spectrum, while the second
represents the largest normalization compatible with the
phenomenological bounds applicable to the stochastic GW
backgrounds. These quantities are of different nature and
in order to be more precise let us consider an example.

Suppose, for simplicity, that we are dealing with a
logarithmic energy spectrum which is a monotonic function
of the present frequency. Suppose, moreover, that the spectrum
decreases sufficiently fast in the infra-red in order
to be compatible both with the pulsar timing bound and with the
CMB anisotropies bounds. Then the most relevant bound will come,
effectively, from Big-Bang nucleosynthesis (BBN) \cite{sch,wal,cop}.
If at BBN too many massless particles are present they would cause
a faster expansion of the Universe. If the expansion would
be too fast, then, the correct abundances of the light
elements ($^4$He, $^3$ He, Li, D) observed in galaxies
could not be reproduced. Thus we
should require that the total number of massless
particles present in the plasma at BBN should not exceed
the energy density stored, at the same epoch, in radiation.
Therefore, in our particular case, we will have that
$\overline{\Omega}^{\,{\rm max}}$ is determined by demanding
that  \cite{sch}
\begin{equation}
h^2_0\,\int\,\Omega_{\rm GW}(f,\eta_0)\;{\rm d}\ln{f}\,<\,0.2\,
h_0^2\,\Omega_{\gamma}(\eta_0)\,\simeq\,5\,\times\,10^{-6},
\label{ns}
\end{equation}
where $\Omega_{\gamma}(\eta_0)\,=\,2.6\,\times\,10^{-5}\,h_0^{-2}$
is the present fraction of critical energy density stored in radiation.
According to our definition, $\overline{\Omega}^{\,{\rm max}}$ is
the maximal normalization of the spectrum compatible with the previous
inequality, namely,
\begin{equation}
h_0^2\,\overline{\Omega}^{\,{\rm max}}\,\simeq\,
\frac{5\,\times\,10^{-6}}{{\cal I}}\;, \qquad
{\cal I}\,= \,\int_{f_{\rm ns}}^{f_{\rm max}}\,
\omega(f)\;{\rm d}\ln{f}.
\label{NSnorm}
\end{equation}
Notice that  $f_{\rm ns}\,\simeq\,10^{-10}$ Hz is the present value
of the frequency corresponding to the horizon at the nucleosynthesis
time; $f_{\rm max}$ stands for  the maximal frequency of the spectrum
and it depends, in general, upon the specific theoretical model.
If the spectrum has different slopes, $\overline{\Omega}^{\,{\rm max}}$
will be determined not only by the nucleosynthesis bound but also by
the combined action of the CMB anisotropy bound \cite{2,ben} and
of the pulsar timing bound \cite{kas}. Indeed, we know that
the very small fractional timing error in the arrival times of
the millisecond pulsar's pulses implies that
$\Omega_{{\rm GW}}\,\laq\,10^{-8}$ for a frequency which is roughly
comparable with the inverse of the observation time along which
pulsars have been monitored
(i.e., $\omega_{\rm p}\,\sim\,1/T_{\rm obs}\,=\,10^{-8}$ Hz). Moreover,
the observations of the large scale anisotropies in the microwave sky
\cite{ben,tur} imply that the graviton contribution to the integrated
Sachs-Wolfe effect has to be smaller than (or at most of the
order of) the detected amount of anisotropy. This observation implies
that $\Omega_{{\rm GW}}\,\leq\,6.9\,\times\,10^{-11}$ for
frequencies ranging between the typical frequency of the present
horizon and a frequency thirty of forty times larger. In the case of
a logarithmic energy density with decreasing slope the
$\overline{\Omega}^{\,{\rm max}}$ will be mainly determined by the
Sachs-Wolfe bound and it will be the maximal normalization of the
spectrum compatible with such a bound.

In general , we will have that
$\overline{\Omega}^{\,{\rm th}}\,\leq\,\overline{\Omega}^{\,{\rm max}}$,
namely the theoretical normalization of the spectrum is bounded, from
above, by the maximal normalization compatible with all the
phenomenological bounds. Therefore, the mismatching
 between these  quantities can be interpreted as an
effective measure of the theoretical error
in the determination of the absolute normalization of the spectrum.

Since $\omega(f)$ enters (in a highly non-linear way) into the form
of $J$ (as defined in Eq. (\ref{Jint})), the corresponding
$\overline{\Omega}$ in Eq. (\ref{sens}) will be different for any
(specific) frequency dependence in $\omega(f)$. The consequence
of this statement is that it is not possible to give a general
(and simple) relation between the sensitivity at a given frequency,
the spectral slope and the (generic) theoretical amplitude of the
spectrum. However, given the form of the theoretical spectrum, the
phenomenological bounds (depending upon the theoretical slope)
will fix uniquely the theoretical error and the maximal achievable
sensitivity. So, if we want to evaluate the performances of the VIRGO
pair we should pick up a given class of theoretical models (characterized
by a specific functional form of $\omega(f)$) and compute
the corresponding sensitivity. The same procedure should then be repeated
for other classes of models and, only at the end, the
respective sensitivities can be compared.

\renewcommand{\theequation}{3.\arabic{equation}}
\setcounter{equation}{0}
\section{Primordial gravitons versus  VIRGO*VIRGO}

We can consider, in principle, logarithmic energy spectra with
hypothetical analytical forms and arbitrary normalizations.
If the logarithmic energy spectrum is either a flat or a decreasing
function of the present frequency \cite{gri}, we can expect,
in general, that the theoretical signal will be of the
order of (but smaller than) $10^{-15}$ \cite{tur,rub} for present
frequencies comparable with the operating window of the
VIRGO pair. This happens because of the combined action of the
Sachs-Wolfe bound together with the spectral behavior of the
infra-red branch of the spectrum produced thanks to the
matter-radiation transition.
Of course this observation holds \cite{rub} for  models where the graviton
production occurs because of the adiabatic variation
of the background geometry \cite{1,gri} \footnote{ An exception to this
 assessment is represented by cosmic strings models
 leading to a flat logarithmic energy spectrum for frequencies
 between $10^{-12}$ Hz and $10^{-8}$ Hz \cite{vac,cal}.
Another possible exception is
 given by the gravitational power radiated by magnetic
 (and hypermagnetic) \cite{gio1}
 knot configurations at the electroweak scale \cite{sha1}.}.

In order to have large signals falling in the operating window of
the VIRGO pair we should have deviations from scale invariance
for frequencies larger than few mHz. Moreover,
these deviations should go in the direction of increasing
logarithmic energy spectra. This is what happens
in the case of quintessential inflationary models \cite{gio2}.
In this case, however, as we discussed in a previous
analysis \cite{noi}, the BBN bound put strong constraints on the
theoretical signal in the operating window of the
VIRGO pair.

Another class of model leading to a large theoretical
signal for frequencies between few Hz and 10 kHz  is
represented by string cosmological models \cite{9,10,11}.
Therefore, in order
to evaluate the performances of the VIRGO pair
and in order to implement a procedure of selective noise
reduction we will use string cosmological spectra.

\subsection{Minimal models of pre-big-bang}

In string cosmology
and, more specifically, in the pre-big-bang scenario, the curvature scale
and the dilaton coupling are both growing in cosmic time. Therefore
the graviton spectra will be {\em increasing} in frequency  instead of
{\em decreasing} as it happens in ordinary inflationary models.

In the context of string cosmological scenarios the Universe
starts its evolution in a very weakly coupled regime with vanishing
curvature and dilaton coupling. After a phase of sudden
growth of the
curvature and of the coupling the corrections to the tree level action
become important and the Universe enters a true stringy phase which is
followed by the ordinary radiation dominated phase. It should be stressed
that the duration of the stringy phase is not precisely known
and it could happen that all the physical scales contained
within our present Hubble radius crossed the horizon during the stringy phase
as pointed out in \cite{gas1}.

The maximal amplified frequency of the graviton spectrum is given
by \cite{9,11}
\begin{equation}
f_{1}(\eta_0)\,\simeq\,64.8\,\sqrt{g_1}\,\left(\,\frac{10^{3}}
{n_r}\,\right)^{1/12}\; {\rm GHz}
\end{equation}
where $n_{r}$ is the effective number of spin degrees
of freedom in thermal equilibrium at the end of the stringy phase,
and $g_1\,=\,M_{s}/\mpl$ where $M_s$ and $\mpl$ are the string
and Planck masses, respectively. Notice that $g_1$ is the value
of the dilaton coupling at the end of the stringy phase, and
is typically of the order of $10^{-2}\,\div\,10^{-1}$ \cite{kap}.
As we can see from the previous equation the dependence upon
$n_r$ is quite weak.
In order to red-shift the maximal amplified frequency of the
spectrum from the time $\eta_1$ (which marks the beginning of
the radiation dominated evolution) up to the present time we
assumed that the cosmological evolution prior to $\eta_0$ and
after $\eta_1$ is adiabatic. Minimal models of pre-big-bang are
the ones where a dilaton dominated phase is followed by a stringy
phase which terminates at the onset of the radiation dominated
evolution. In the context of minimal models, the function
$\omega (f)$ introduced in Eq. (\ref{1}) can be written as
\begin{equation}
\omega (f)\,=\,
\left\{
\begin{array}{lc}
\dis z_s^{- 2 \beta}\,\left(\,\frac{f}{f_s}\,\right)^3\,\left[\,1\,+\,
z_s^{2 \beta - 3}\,-\,\frac12\,\ln{\frac{f}{f_s}}\,\right]^2
& \dis  \qquad f\,\le\,f_s\,=\,\frac{f_1}{z_s} \\ [15pt]
\dis \left[\,\biggl(\frac{f}{f_1}\biggr)^{3 - \beta}\,+\,
\biggl(\frac{f}{f_1}\biggr)^{\beta}\,\right]^2
& \qquad  f_s\,<\,f\,\le\,f_1
\end{array}
\right.
\label{minth}
\end{equation}
where,
\begin{equation}
\dis \beta\,=\,\frac{\ln\,(g_1/g_s)}{\ln\,z_s}.
\end{equation}
In this formula $z_s\,=\,f_1/f_s$ and $g_s$ are, respectively, the
red-shift during the string phase
and the value of the coupling constant at the end of the dilaton dominated
phase.
The first of the two branches appearing in Eq. (\ref{minth})
is originated by modes leaving the horizon
during the dilaton dominated phase and re-entering
during the radiation dominated phase. The second branch is
mainly originated by modes leaving the horizon during the
stringy phase and re-entering always in the radiation
dominated phase. The theoretical normalization \cite{9,11}
\begin{equation}
\overline{\Omega}^{\,\rm th}\,=\,2.6\,g_1^2\,\left(\,\frac{10^3}{n_r}\,
\right)^{1/3}\,\Omega_{\gamma}(\eta_0)\;,
\label{omth}
\end{equation}
multiplied by  $\omega(f)$ (as given in Eq. (\ref{minth})) leads
to the theoretical form of the spectrum.
Notice that $n_r$ is of the order of $10^2\,\div\,10^{3}$ (depending
upon the specific string model) and it represents a theoretical
uncertainty.

However, as anticipated in the previous section, the theoretical
normalization of the spectrum should be contrasted with the one
saturating the BBN bound (i.e., $\overline{\Omega}^{\,{\rm max}}$).
This quantity is obtained by Eq. (\ref{NSnorm}), where in the case
under consideration
\begin{equation}
{\cal I}\,=\,{\cal I}_{d}\,+\,{\cal I}_{s} \qquad {\rm with}
\qquad {\cal I}_{d}\,=\,\int_{f_{\rm ns}}^{f_s}\,\frac{{\rm d}f}{f}\,
\omega (f)\;,\qquad {\cal I}_{s}\,=\,\int_{f_{s}}^{f_1}\,
\frac{{\rm d}f}{f}\,\omega (f)\;.
\end{equation}
In the case of minimal pre-big-bang models
the analytical expressions of ${\cal I}_{d}$ and ${\cal I}_{s}$
are given by
\begin{eqnarray}
{\cal I}_{d} &=& z_{s}^{- 2 \beta}\,\left\{\,\frac{1}{54}\,
(z_s^2 + 6\,z_s + 18)\,-\,\frac{1}{108}\,\left(\,\frac{f_{\rm ns}}{f_{s}}\,
\right)^3\,\left[\,2\,(z_s^2 + 6\,z_s + 18) \right. \right.
\nonumber \\
& & \qquad \qquad \qquad \left. \left.
-\,6\,z_s\,(z_s + 6)\,\ln{\frac{f_{\rm ns}}{f_s}}\,+
\,9\,z_s^2\,\ln^2{\frac{f_{ns}}{f_s}}\,\right]\,
\right\}\;,\nonumber\\[10pt]
{\cal I}_{s} &=& \frac{3}{2\,\beta\,(3 - 2 \beta)}\,+\,
\frac{z_s^{2 \beta - 6}}{2 \beta - 6}\,-\,\frac{z_s^{- 2 \beta}}
{2 \beta}\;.
\end{eqnarray}
In the case of non-minimal models of pre-big-bang
the integrals determining
the BBN bound are instead
\begin{eqnarray}
{\cal I}_{1} &=& A(\sigma, z_s)\,+\,B(\sigma, z_s)\,\ln{z_s}\,+\,
C(\sigma, z_s)\,\ln^2{z_s}\;,\nonumber\\[10pt]
{\cal I}_{2} &=& \frac{z_s^{-4}}{4}\,\left(\,z_s^{\sigma - 2}\,+\,
z_s^{2 + \sigma}\,\right)\,\left(\,z_s^{- 4}\,-\,z_r^{- 4}\,\right)\,
(1\,+\,\ln{z_s})^2\;,
\end{eqnarray}
where and $z_r\,=\,f_1/f_r$ and
\begin{eqnarray}
A(\sigma, z_s) &=& - \frac{z_s^{2 \sigma}}{16\,(\sigma^2 - 4)^3}\,
\left\{\,13\,z_s^{- 2(2 + \sigma)}\,(\sigma^2 - 4)^3 \right.
\nonumber \\
& & \qquad \qquad \qquad \qquad \left. -\,4\,z_s^{-4}\,
(\sigma + 2)^3\,(2 \sigma^2 - 10 \sigma + 13) \right.
\nonumber\\
& & \qquad \qquad \qquad \qquad \left. 
+\,4\,z_s^{- 4( 1 + \sigma)}\,(\sigma - 2)^3\,
(2 \sigma^2 + 10 \sigma + 13)\right.
\nonumber\\
& & \qquad \qquad \qquad \qquad \left. 
-\,z_s^{- 2 \sigma}\,
(13 \sigma^6 - 172 \sigma^4 + 832 \sigma^2 - 1664)\,
\right\}\;, \nonumber \\[10pt]
B(\sigma, z_s) &=& \frac{z_s^{2 \sigma - 4}}{4\,(\sigma^2 - 4)^2}\,
\left\{\,2\,(\sigma + 2)^2\,(2 \sigma - 5)\,-\,2\,z_s^{- 4 \sigma}\,
(\sigma - 2)^2\,(2 \sigma + 5)\right.
\nonumber\\
& & \qquad \qquad \qquad \qquad \left. 
-\,5\,z_s^{2 \sigma}\,(\sigma^2 - 4)^2\,\right\}\;,
\nonumber\\[10pt]
C(\sigma, z_s) &=& \frac{z_s^{4 - 2 \sigma}}{2\,(\sigma^2 - 4)}\,
\left\{\,2\,-\,z_s^{- 4 \sigma}\,(\sigma - 2)\,+\,\sigma\,
z_s^{- 2 \sigma}\,(\sigma^2 - 4))\right\}\;.
\end{eqnarray}

In the intermediate
frequency region of the graviton spectra an important bound comes
from the pulsar timing measurements. Therefore, if one ought to
consider rather long stringy phases (i.e., large $z_s$), the BBN
constraint should be supplemented by the requirement that
$\Omega_{\rm GW}(10^{-8}\,{\rm Hz})\,<\,10^{-8}$ \cite{kas}. We will come
back to this point later.

Following the explicit expression of the function $\omega (f)$,
Eq. (\ref{sens}) can be re-written as follows:
\begin{equation}
h_0^2\,\overline{\Omega}\,\simeq\,4\,\times\,10^{-7}\,\left(\,
\frac{1\,{\rm yr}}{T}\,\right)^{1/2}\,
\frac{{\rm SNR}^2}{\sqrt{J_d^2\,+\,J_s^2}},
\end{equation}
where, introduced the following notation
\begin{eqnarray}
J_{k} &=& \int_{\nu_{\rm m}}^{\nu_{s}}\,{\rm d}\nu\,
\frac{\gamma^2 (f_0 \nu)}{\Sigma_{n}^{(1)} (f_0\nu)\,
\Sigma_{n}^{(2)} (f_0\nu)}\,\ln^{k}{\nu}\;,
\qquad k\,=\,0,1,2,3,4 \nonumber\\[10pt]
J_{\pm m (3 - 2 \beta)} &=& \int_{\nu_s}^{\nu_{\rm M}}\,
{\rm d}\nu\,\frac{\gamma^2 (f_0 \nu)}{\Sigma_{n}^{(1)} (f_0 \nu)\,
\Sigma_{n}^{(2)} (f_0\nu)}\,\nu^{\pm m (3 - 2 \beta)}\;,
\;\;m\,=\,1,2 \\[10pt]
C_d &=& 1\,+\,z_s^{2 \beta - 3}\,+\,\frac12\,\ln{\nu_s}\;, \nonumber
\end{eqnarray}
one has
\begin{eqnarray}
J_{d} &=& \frac{z_s^{3 - 2 \beta}}{\nu_1^3}\,
\left(\,C_d^4 J_0\,-\,2 C_{d}^3 J_1\,+\,\frac32 C_{d}^2 J_2\,-\,
\frac12 C_{d} J_3\,+\,\frac{1}{16} J_{4}\,\right)^{1/2}\;,
\nonumber \\[10pt]
J_{s} &=& \frac{1}{\nu_1^3}\,\left(\,6 J_0\,+\,
\frac{J_{6 - 4 \beta}}{\nu_1^{6 - 4 \beta}}\,+\,
\frac{J_{4 \beta - 6}}{\nu_1^{4 \beta - 6}}\,+\,
4 \frac{J_{3 - 2 \beta}}{\nu_1^{3 - 2 \beta}}\,+\,
4 \frac{J_{2 \beta - 3}}{\nu_1^{2 \beta - 3}}\,\right)^{1/2}\;.
\end{eqnarray}

The previous expressions are general in the sense that they are
applicable for a generic value of $f_s$.
If $f_{\rm m}\,<\,f_{s}\,<\,f_{\rm M}$ then both $J_{s}$ and $J_{d}$
give contribution to the sensitivity. If, on the other hand
$f_s\,<\,f_{\rm m}$ (i.e., a long stringy phase) the main contribution
to the sensitivity will come from $J_{s}$. The integrals appearing in
$J_{d,s}$ have to be evaluated numerically. In all our calculations we
will assume that both VIRGO detectors are characterized by the same
(rescaled) NPS (reported in Eq. (\ref{snrrescaled})).

The main steps of our calculation are the following. We firstly fix $g_1$ and
for each pair $(z_s,\,g_1/g_s)$ (within the range of their physical
value) we compute $\overline{\Omega}$  (for $T\,=\,1$ yr and SNR = 1),
and $\overline{\Omega}^{\,{\rm max}}$. We then compare these two quantities
to the theoretical normalization given in Eq. (\ref{omth}). If
$\overline{\Omega}^{\,{\rm th}}$ will be larger than $\overline{\Omega}$
(but smaller than $\overline{\Omega}^{\,{\rm max}}$)
we will say that the theoretical
signal will be ``visible'' by the VIRGO pair. In this way we will
identify in the plane $(z_s,\,g_1/g_s)$ a visibility region according
to the sensitivity of the VIRGO pair. The theoretical error on the border
of this region can be estimated by substituting
$\overline{\Omega}^{\,{\rm max}}$ to $\overline{\Omega}^{\,{\rm th}}$.

To illustrate this point we consider a specific case. The value of the
coupling at the end of the stringy phase can be estimated to lie between
0.3 and 0.03 \cite{kap}.
The knowledge of $g_1$ will not fix uniquely the theoretical
spectrum which does also depend on the number of relativistic degrees of
freedom at the end of the stringy phase. Therefore, the theoretical error
in the determination of the absolute normalization of the spectrum could
be also viewed as the error affecting the determination of $n_r$.
In all the plots shown we will take, when not otherwise stated,
$g_1\,=\,1/20$ and $n_{r}\,=\,10^3$ as fiducial values. Different choices
of $g_1$ will lead to similar results. We will also
assume that the overlap reduction function associated with the
pair is the one reported in the curve A of Fig. \ref{over}.
\begin{figure}[!hb]
\begin{center}
\begin{tabular}{cc}
      \hbox{\epsfxsize = 6.2 cm  \epsffile{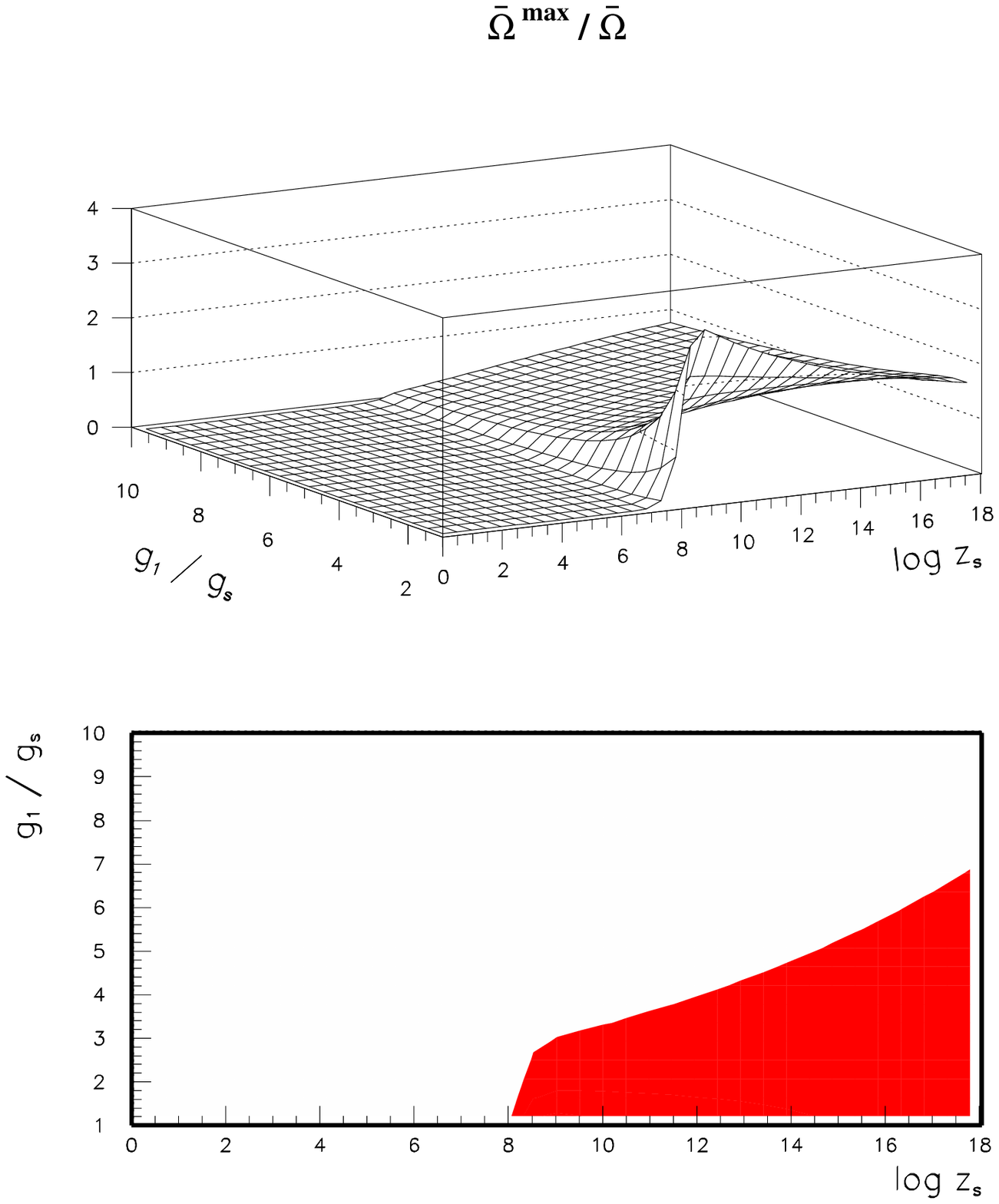}} &
      \hbox{\epsfxsize = 6.2 cm  \epsffile{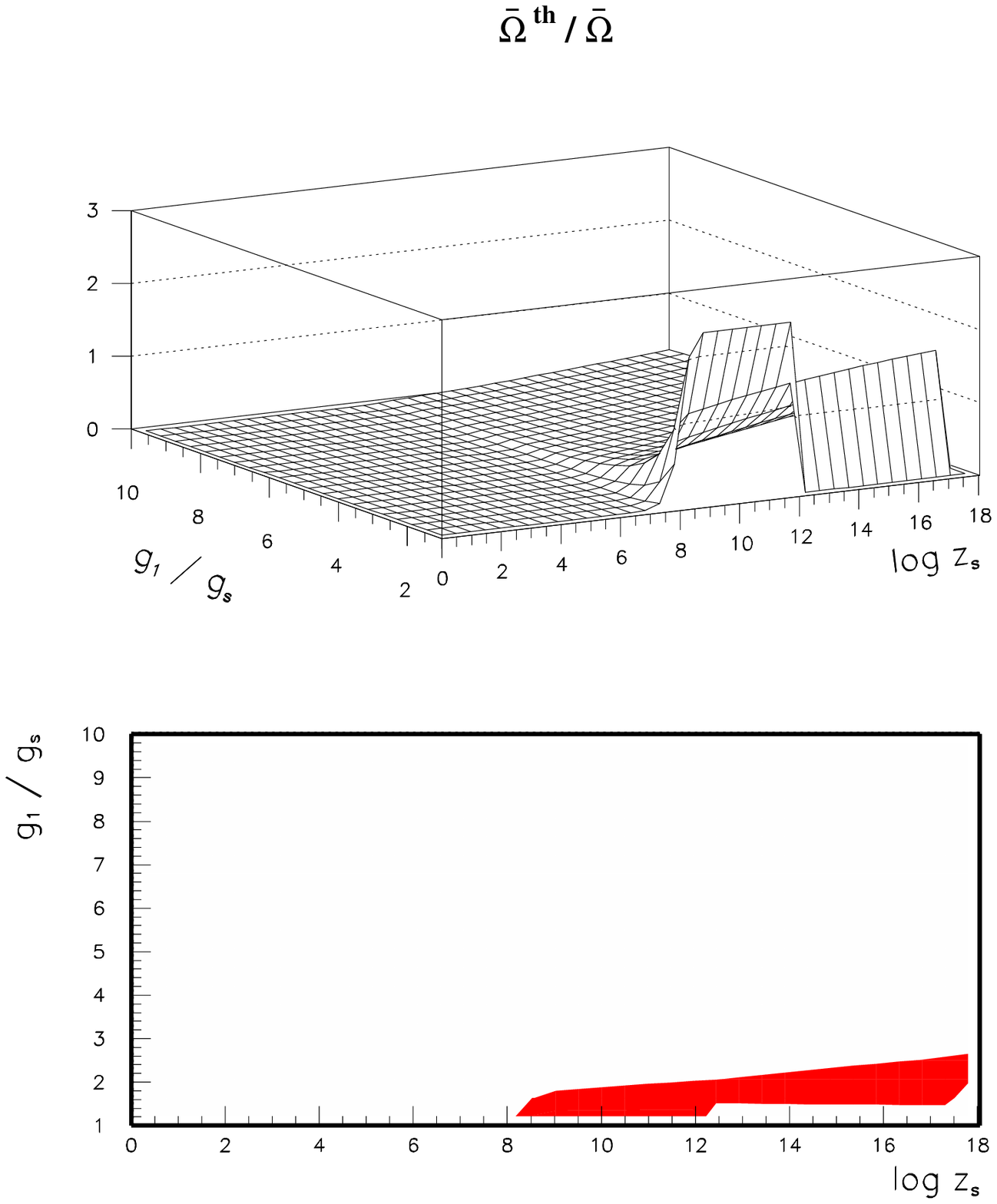}} \\
\end{tabular}
\end{center}
\vspace*{-1.5cm}
\caption[a]{We report the ratios
$\overline{\Omega}^{\,{\rm max}}/\overline{\Omega}$ (left) and
$\overline{\Omega}^{\,{\rm th}}/\overline{\Omega}$ (right)
as a function of $g_1/g_s$ and $\log{z_s}$  ($\overline{\Omega}$
is calculated for $T\,=\,1$ yr and SNR = 1). The lower contour plots
show the regions where these ratios are greater than 1.
The shaded area (bottom right) represents the region where the combination
of the theoretical parameters is such that the corresponding
$\overline{\Omega}^{\,{\rm th}}$ does not violate the BBN bound.
As we can see the visibility region is reduced.
The difference between the shaded area in the right plot and the one
in the left plot measures the error made by assuming as normalization of
the spectrum not the theoretical one but the maximal one compatible
with the BBN. The value $z_{s}\,=\,10^{8}$ roughly corresponds to
$f_s\,\sim\,f_0$.
}
\label{minrat}
\end{figure}

In Fig. \ref{minrat} (top left) we report the result of our
calculation for the ratio between
$\overline{\Omega}^{\,{\rm max}}$ and $\overline{\Omega}$ as a
function of $g_1/g_s$ and $\log{z_s}$. The contour plot (bottom left)
shows the region of the plane $(\log{z_s},\,g_1/g_s)$
where this ratio is greater than 1, i.e. the maximal visibility
region allowed by the BBN bound. In the opposite case, i.e.,
$\overline{\Omega}^{\,{\rm max}}/\overline{\Omega}\,<\,1$,
the VIRGO pair is sensitive to a region excluded by the BBN.
In the right part of Fig. \ref{minrat} we go one step further and
we plot the ratio between $\overline{\Omega}^{\,{\rm th}}$ and
$\overline{\Omega}$. The shaded area in the
contour plot (bottom right) is the region of the plane $(\log{z_s},\,g_1/g_s)$
where the conditions
$\overline{\Omega}^{\,{\rm th}}/\overline{\Omega}\,>\,1$ and
$\overline{\Omega}^{\,{\rm max}}/\overline{\Omega}\,>\,1$ are
simultaneously met. The shaded area in this plot defines the visibility
region of the VIRGO pair {\em assuming} the theoretical normalization
of the spectrum. From  Fig. \ref{minrat},
by ideally subtracting the shaded area of the left contour
plot from the shaded area of the right contour plot
we obtain an estimate of the theoretical error.
The results we just presented can be obviously recovered for different
values of $g_1$ close to one. However, if $g_1$ gets too small
(and typically below 1/25) the visibility area gets smaller and
smaller eventually disappearing.

The visibility regions appearing in Fig. \ref{minrat} extend from
intermediate values of $z_s$ (of the order of $10^{8}$) towards
large values of $z_s$ (of the order of $10^{18}$). Notice that
for our choice of $g_1$, $f_s$ can become as small as $10^{-8}$
for $z_s $ of the order of $10^{18}$. As we recalled in the
previous Section, this frequency corresponds to the inverse of
the observation time along which pulsar signals have been monitored
and, therefore, for this  frequency, a further ``local'' bound
applies to the logarithmic energy spectra of relic gravitons.
This bound implies that
$\Omega_{\rm GW}(10^{-8}\,{\rm Hz})\,<\,10^{-8}$. In our examples,
the compatibility with the BBN bound implies also that the
pulsar timing constraint is satisfied. Given our choice
for $g_1$ we can clearly see that the visibility regions depicted
in Fig. \ref{minrat} extend for values of $g_s$ which can be as
small as 1/160 (or as small as 1/60 in the case of right part of
Fig. \ref{minrat}).

\subsection{Non-minimal models of pre-big-bang}

In the context of minimal models of pre-big-bang, the
end of the stringy phase coincides with the
onset of the radiation dominated evolution.
At the moment of the transition to the
radiation dominated phase the dilaton sits
at its constant value. This
means that $g_1\,\sim\,0.03\,\div\,0.3$ at the beginning
of the radiation dominated evolution.
As pointed out in \cite{gas1}, it is  not be impossible to imagine
a scenario where the coupling constant is still
growing while the curvature scale starts decreasing in time.
In this type of scenario the stringy phase is followed
by a phase where the dilaton still increases, or, in other
words, the coupling constant is rather small at the moment
where the curvature starts decreasing so that $g_1\,\ll\,1$.

After a transient period (whose precise duration will be fixed by the
value of $g_1$), we will have that the radiation dominated evolution
will take place when the value of the coupling constant will be of
order one (i.e., $g_r\,\sim\,1$).
An interesting feature of this speculation is that the
graviton spectra will not necessarily be monotonic \cite{gas1}
(as the ones considered in the previous  analysis).
We then find interesting to apply our considerations also to this case.

The function $\omega (f)$ in the non-minimal model described
above is given by \cite{gas1}
\bite
\item $ \dis f_r\,<\,f\,\le\,f_s \,=\,\frac{f_1}{z_s} $
$$ \dis
\omega (f)\,=\,
\left(\,\frac{g_r}{g_1}\,\right)^{2/\sqrt{3}}\,
\left(\,\frac{f}{f_1}\,\right)^4\,
\left[\,\left(\,\frac{f_s}{f_1}\,\right)^{- \sigma}\,+\,
\left(\,\frac{f_s}{f_1}\,\right)^{\sigma}\,\right]^2\,
\left(\,1\,-\,\ln{\frac{f_s}{f_1}}\,\right)^2
$$
\item $ \dis f_s\,<\,f\,\le\,f_1 $
\begin{equation}
\omega (f)\,=\,
\left(\,\frac{g_r}{g_1}\,\right)^{2/\sqrt{3}}\,
\left[\,\left(\,\frac{f}{f_1}\,\right)^{2 - \sigma}\,+\,
\left(\,\frac{f}{f_1}\,\right)^{2 + \sigma}\,\right]^2\,
\left(\,1\,-\,\ln{\frac{f}{f_1}}\,\right)^2
\label{nonminth}
\end{equation}
\eite
where, in the present case
\begin{equation}
f_1\,\simeq\,64.8\,\sqrt{g_1}\,\left(\,\frac{g_r}{g_1}\,
\right)^{1/2\sqrt{3}}\,\left(\,\frac{10^3}{n_r}\,\right)^{1/12}
\;{\rm GHz}\;, \qquad f_r\,=\,\left(\,\frac{g_r}{g_1}\,
\right)^{- 2/\sqrt{3}}\,f_1\;.
\end{equation}
(Notice that the form of $\omega(f)$ reported in
\cite{gas1} differs from our expression only by
logarithmic correction whose presence is, indeed, not
relevant.)
The frequency $f_r$ corresponds to the
onset of the radiation dominated evolution.
If we adopt a purely phenomenological approach we can
say that $f_r$ has to be bounded (from below)
since we want the Universe to be
radiation dominated not later than the BBN
epoch. Thence, we have that $f_r\,>\,f_{\rm ns}$. Recalling
the value of the nucleosynthesis frequency and
assuming that $g_r\,\simeq\,1$ this condition
implies $g_1\,\gaq\,8.2\,\times\,10^{-16}$. This simply
means that in order not to conflict with
the correct abundances of the light elements we have to
require that the coupling constant should not be too small
when the curvature starts decreasing. Notice that
for frequencies $f\,<\,f_r$ the spectrum evolves
as $f^{-3}$.
The ultra-violet branch of the spectrum is mainly
originated by modes leaving the
horizon during the stringy phase and re-entering when
the dilaton coupling is still increasing.

Concerning the non-minimal spectra few comments are in order.
Owing to the fact that $g_1$ can be as small as
$10^{-15}$ we have that the highest frequency of the spectrum
can become substantially smaller than in the
minimal case. Moreover, the spectrum might also be
non-monotonic with a peak at $f_s$. Looking at the
analytical form of the spectrum we see that
this behavior occurs if $\sigma\,>\,2$.  We remind that
$\sigma $ parametrizes the spectral slope in the phase where
the curvature scale decreases but the dilaton coupling
is still growing \cite{gas1}.
A non-monotonic logarithmic energy spectrum
(with a maximum falling in the sensitivity region of
the VIRGO pair) represents an interesting possibility.

The results of our calculation for $g_1\,=\,10^{-12}$,
 $n_r\,=\,10^3$, $g_r\,=\,1$, and $\sigma\,>\,2$ are reported
in Fig. \ref{nminrat}.
\begin{figure}[!ht]
\begin{center}
\begin{tabular}{cc}
      \hbox{\epsfxsize = 6.2 cm  \epsffile{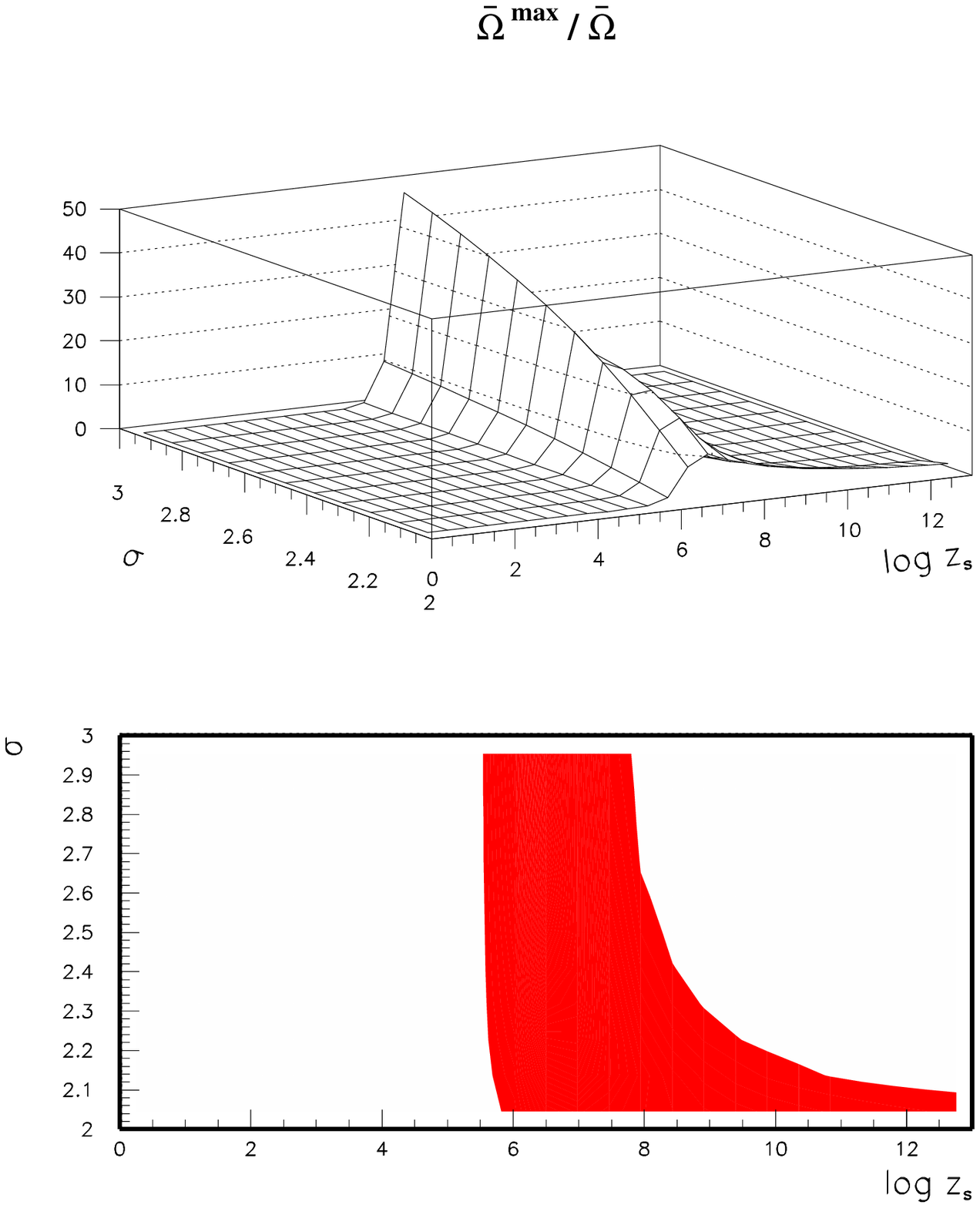}} &
      \hbox{\epsfxsize = 6.2 cm  \epsffile{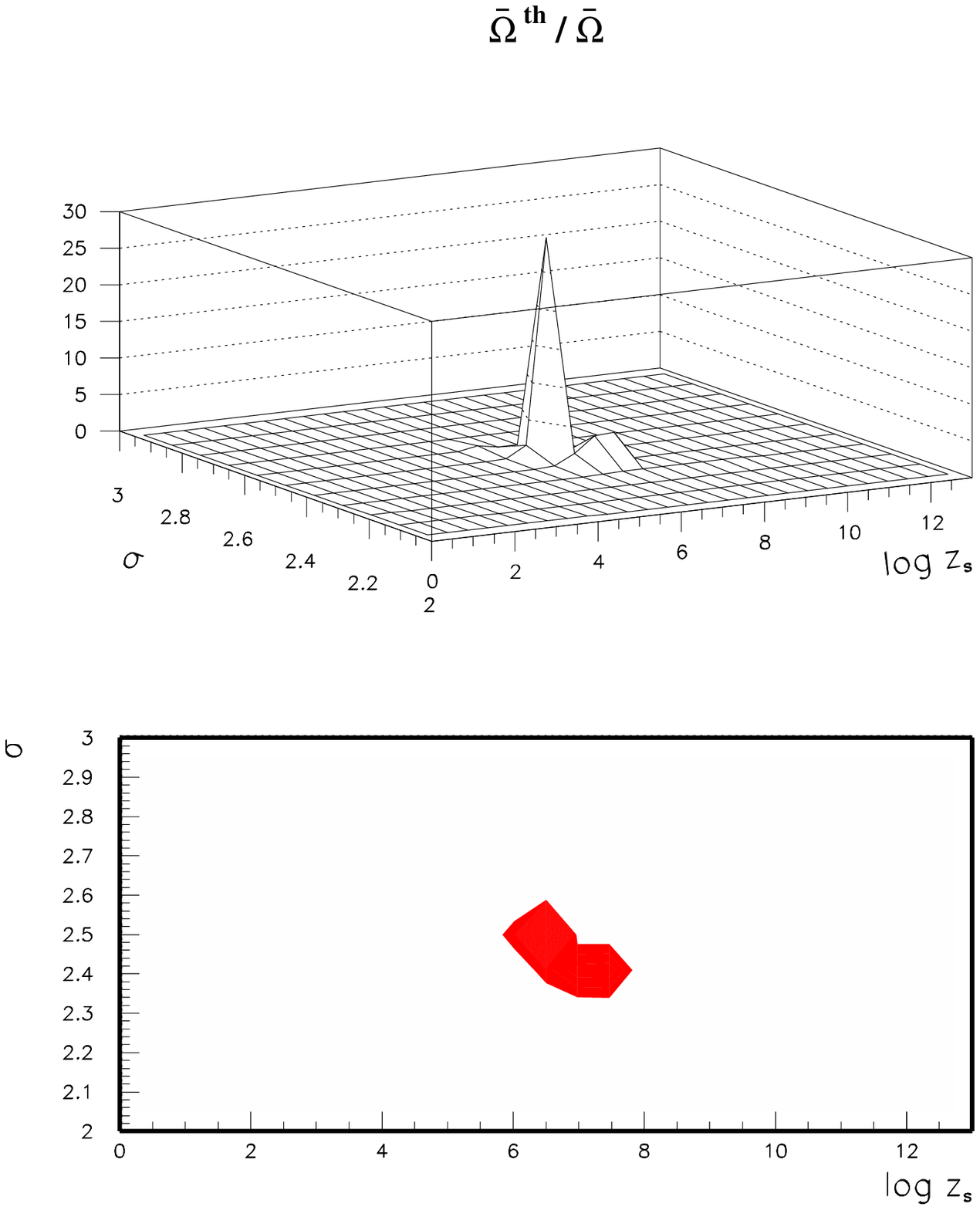}} \\
\end{tabular}
\end{center}
\vspace*{-1.5cm}
\caption[a]{In order to make clear the comparison with the
visibility region of the minimal models, we report
$\overline{\Omega}^{\,{\rm max}}/\overline{\Omega}$ (left) and
$\overline{\Omega}^{\,{\rm th}}/\overline{\Omega}$ (right)
as a function of $\sigma$ and of the $\log{z_s}$ in the non-minimal
scenario. Notice that we took $g_1\,=\,10^{-12}$, $n_r\,=\,10^3$,
and $g_r\,=\,1$. As for Fig. \ref{minrat}, the shaded areas in the
lower contour plots represent the region where each ratio is greater
than 1, and, in the case of the right plot, also the BBN is satisfied.}
\label{nminrat}
\end{figure}
As done in the case of minimal spectra
we analyse the visibility window in the plane of the relevant
parameters of the model. As we can see from the  left part of
Fig. \ref{nminrat} the region compatible with the BBN is rather
large but it shrinks when we impose the theoretical normalization
( right part of Fig. \ref{nminrat}) which is always
smaller than the maximal normalization allowed by BBN.

It is interesting to compare directly the three dimensional plots appearing
in Fig. \ref{minrat} with the corresponding three dimensional plots
of Fig. \ref{nminrat}. Notice that the shaded region in
the case of minimal models corresponds to ratios
$\overline{\Omega}^{\,\rm max}/\overline{\Omega}$ and
$\overline{\Omega}^{\,\rm th}/\overline{\Omega}$ which can be
 3 or 2, respectively. On the other hand the shaded region in
the case of Fig. \ref{nminrat} corresponds to ratios
$\overline{\Omega}^{\rm max}/\overline{\Omega}$ and
$\overline{\Omega}^{\rm th}/\overline{\Omega}$ which can be,
respectively, as large as 50 or 25. So, in the latter case the
signal is larger for a smaller region of the parameter space.

\renewcommand{\theequation}{4.\arabic{equation}}
\setcounter{equation}{0}
\section{Noise reduction and the visibility region of a VIRGO pair}

There are two ways of looking at the calculations reported in this
paper. One can look at these ideas from a purely theoretical
perspective. In this respect we presented a study of the
sensitivity of a pair of VIRGO detectors to string cosmological
gravitons. There is also a second way of looking at our exercise.
Let us take at face value the results we obtained and let us ask
in what way we can enlarge the visibility region of the VIRGO pair.
In this type of approach the specific form of graviton spectrum is
not strictly essential. We could use, in principle, any motivated
theoretical spectrum. As we stressed, we will use string
cosmological spectra because, on one hand, they are, in our opinion,
 theoretically
motivated and, on the other hand, they give us a signal which
could be, in principle detected. Of course, there are other well
motivated spectra (like the ones provided by ordinary inflationary
models). However, the signal would be, to begin with, quite small.

In this Section we will then consider the following problem.
Given a pair of VIRGO detectors, we suppose to be able,
by some means, to reduce, in a selective fashion,
the contribution of a specific noise source to the detectors
output. The question we ought to address is how the visibility
regions will be modified with respect to the case in which the
selective noise reduction is not present.
We will study the problem for the pair of VIRGO detectors
considered in the previous Sections, i.e., for identical detectors
with NPS given in Eq. (\ref{NPS}), and characterized by the
overlap reduction function of the case A of Fig. \ref{over}.
As for the theoretical graviton spectrum we will focus our
attention on the case of minimal models considered in Section III.A,
 with the same parameters used to produce Fig. \ref{minrat}.
Also here, the quantity $\overline{\Omega}$ will be computed
for $T\,=\,1$ yr and SNR = 1.

As shown in Section II the NPS is characterized by three
dimension-less numbers $\Sigma_{1,2,3}$, and two frequencies
$f_a$ and $f_{b}$. Roughly, $\Sigma_1$ and $\Sigma_2$
control, respectively, the strength of the  pendulum
and pendulum's internal modes noise, whereas $\Sigma_3$ is
related to the shot noise  (see Ref. \cite{sau} for an
accurate description of the phenomena responsible of these
noises). Below the frequency $f_b$ the NPS is dominated by
the seismic noise (assumed to be infinity in Eq. (\ref{NPS})).
The frequency $f_a$ is, roughly, where the NPS gets its minimum.
The frequency behavior of this three contributions and of the
total NPS is shown in Fig. \ref{noise}. The stochastic processes
associated with each source of noise are assumed to be Gaussian
and stationary.
\begin{figure}[!hb]
\centerline{\epsfxsize = 6.2 cm  \epsffile{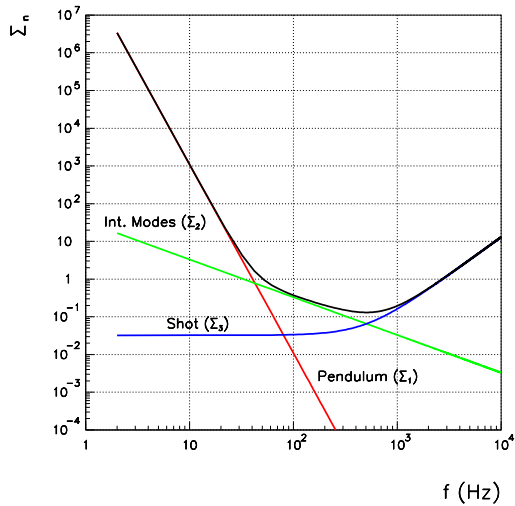}}
\vspace*{-1.0cm}
\caption[a]{The analytical fit of the rescaled noise power
spectrum $\Sigma_n$ defined in Eq. (\ref{NPS}) in the case
of the VIRGO detector. With the full (thick) line we denote
the total NPS. We also report the separated contribution of the three
main (Gaussian and stationary) sources of noise.}
\label{noise}
\end{figure}

In the following, without entering in details concerning the actual
experimental  strategy adopted for the noise reduction, we will
suppose to be able to reduce each of the coefficients $\Sigma_i$ by
keeping the other fixed. In order to make our notation simpler we
define a ``reduction vector''
\begin{equation}
\vec{\rho}\,=\,(\rho_1, \rho_2, \rho_3)\;,
\end{equation}
whose components define
the reduction of the pendulum ($\rho_1$),
pendulum's internal modes ($\rho_2$)
 and shot ($\rho_3$) noises with respect to their fiducial values
appearing in Eq. (\ref{NPS})  (corresponding to the case
$\vec{\rho}\,=\,(1, 1, 1)$).

As shown in Fig. \ref{noise} the pendulum noise dominates the
sensitivity of the detectors in the low frequency region, namely
below about 40 Hz. In Fig. \ref{noired1} we report the results of
our calculation for the case $\vec{\rho}\,=\,(0.1, 1, 1)$.
Here the parameters of the theoretical spectrum are exactly the
same as in Fig. \ref{minrat}. The only change is given by a reduction
of the pendulum noise. From the comparison between Fig. \ref{noired1}
and Fig. \ref{minrat}, we see that the visibility region in the
parameter space of our model gets immediately larger especially
towards the region of small $g_s$. This enlargement is quite
interesting especially in terms of
$\overline{\Omega}^{\,{\rm th}}/\overline{\Omega}$.
\begin{figure}[!hb]
\begin{center}
\begin{tabular}{cc}
      \hbox{\epsfxsize = 6.2 cm  \epsffile{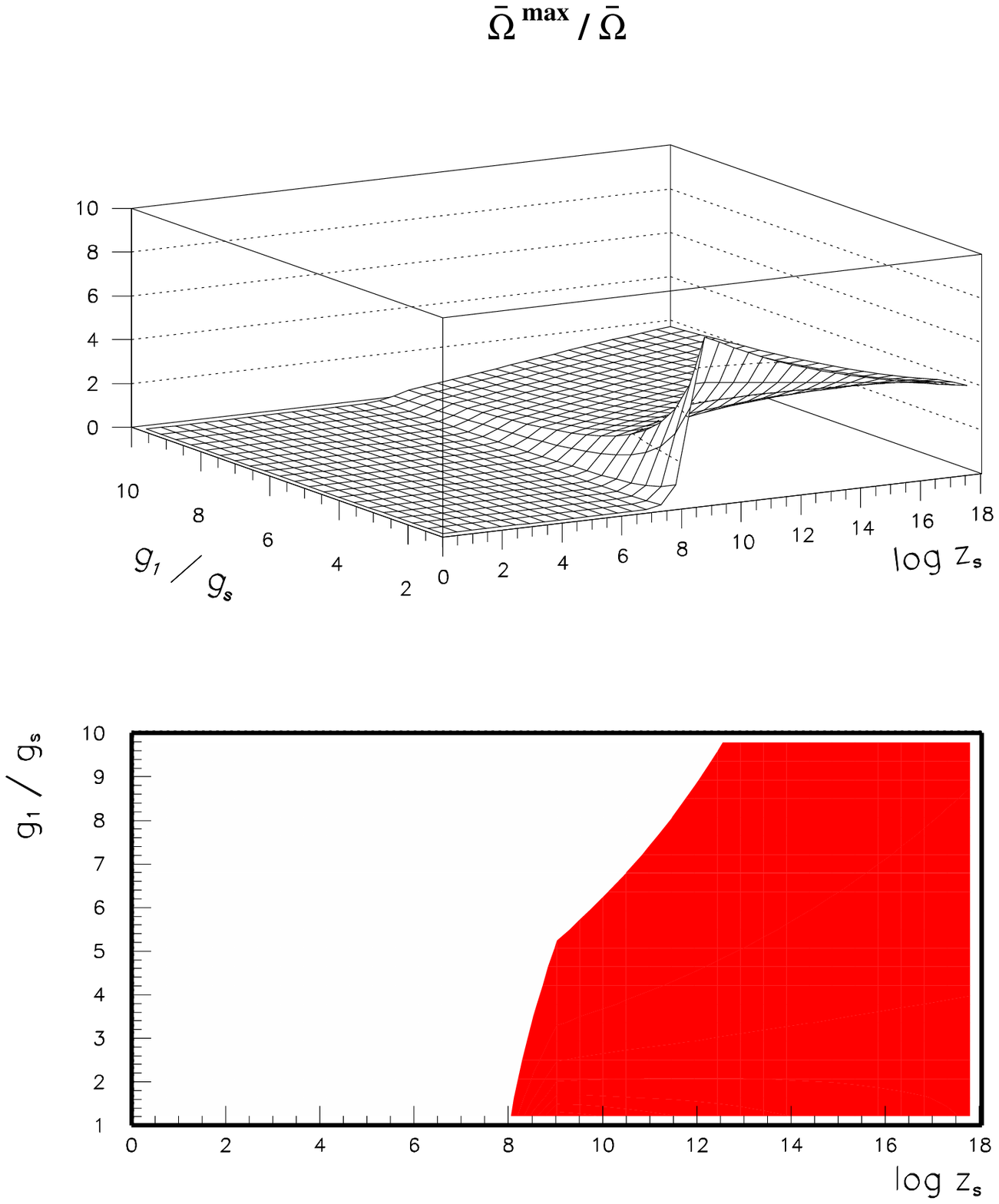}} &
      \hbox{\epsfxsize = 6.2 cm  \epsffile{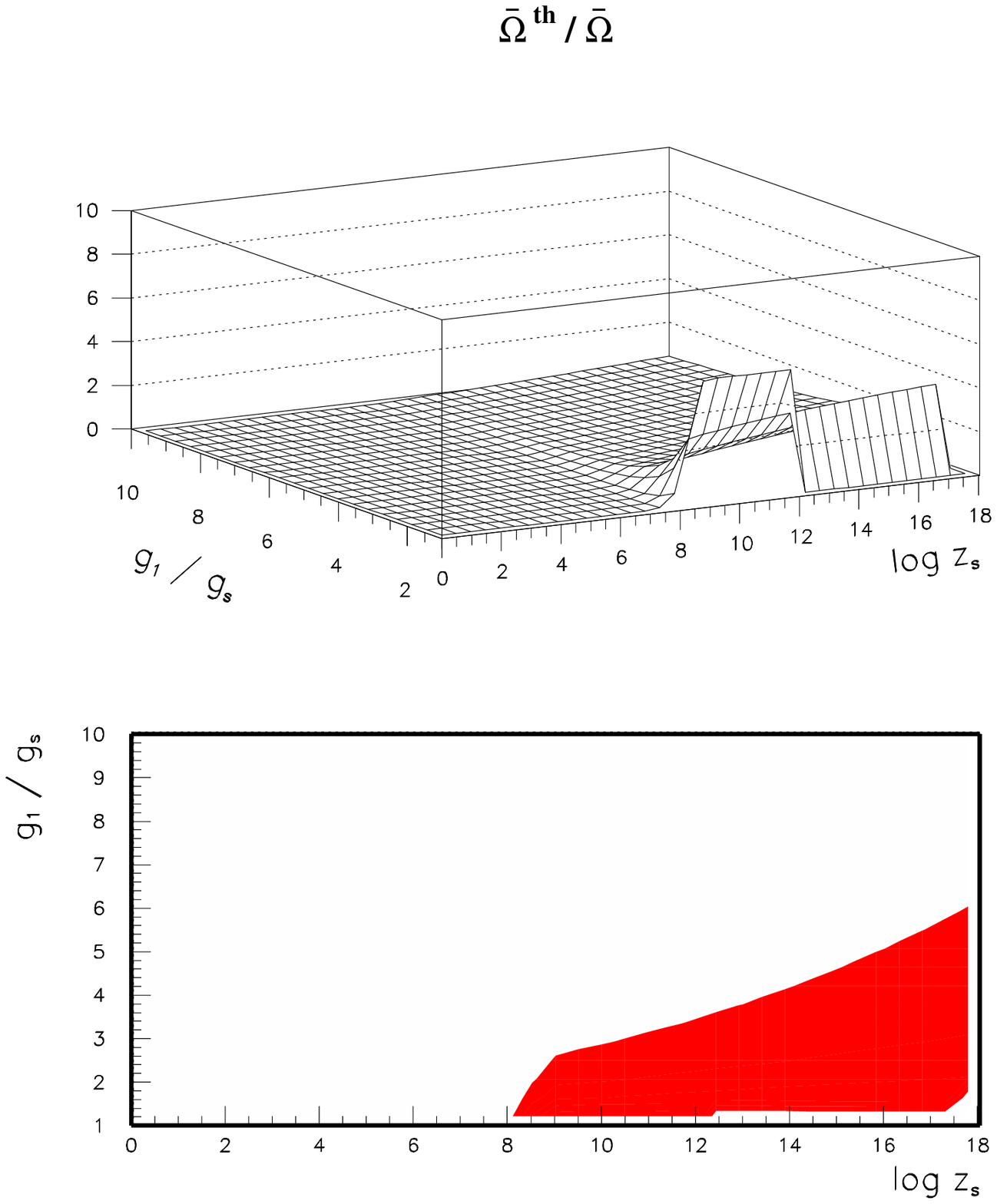}} \\
\end{tabular}
\end{center}
\vspace*{-1.5cm}
\caption[a]{We report the ratios
$\overline{\Omega}^{\,{\rm max}}/\overline{\Omega}$ (left plots), and
$\overline{\Omega}^{\,{\rm th}}/\overline{\Omega}$ (right plots)
in the case in which the shot noise and the noise related to the
pendulum's internal modes are not reduced, whereas the pendulum noise
is diminished by a factor of ten with respect to the values quoted in
Eq. (\ref{NPS}), i.e., $\vec{\rho}\,=\,(0.1, 1, 1)$.}
\label{noired1}
\end{figure}

In the frequency region between 50 and 500 Hz the performances of the
detectors are, essentially, limited by the  pendulum's internal
modes noise. The results obtained for a selective reduction of this
component are summarized in Fig. \ref{noired2},  where the
pendulum and shot noises are left unchanged but the internal modes
component is reduced by a factor of ten. As we
can see the visibility region gets larger and the increase in the area
is comparable with the one obtained by selecting only the  pendulum
noise.
\begin{figure}[!ht]
\begin{center}
\begin{tabular}{cc}
      \hbox{\epsfxsize = 6.2 cm  \epsffile{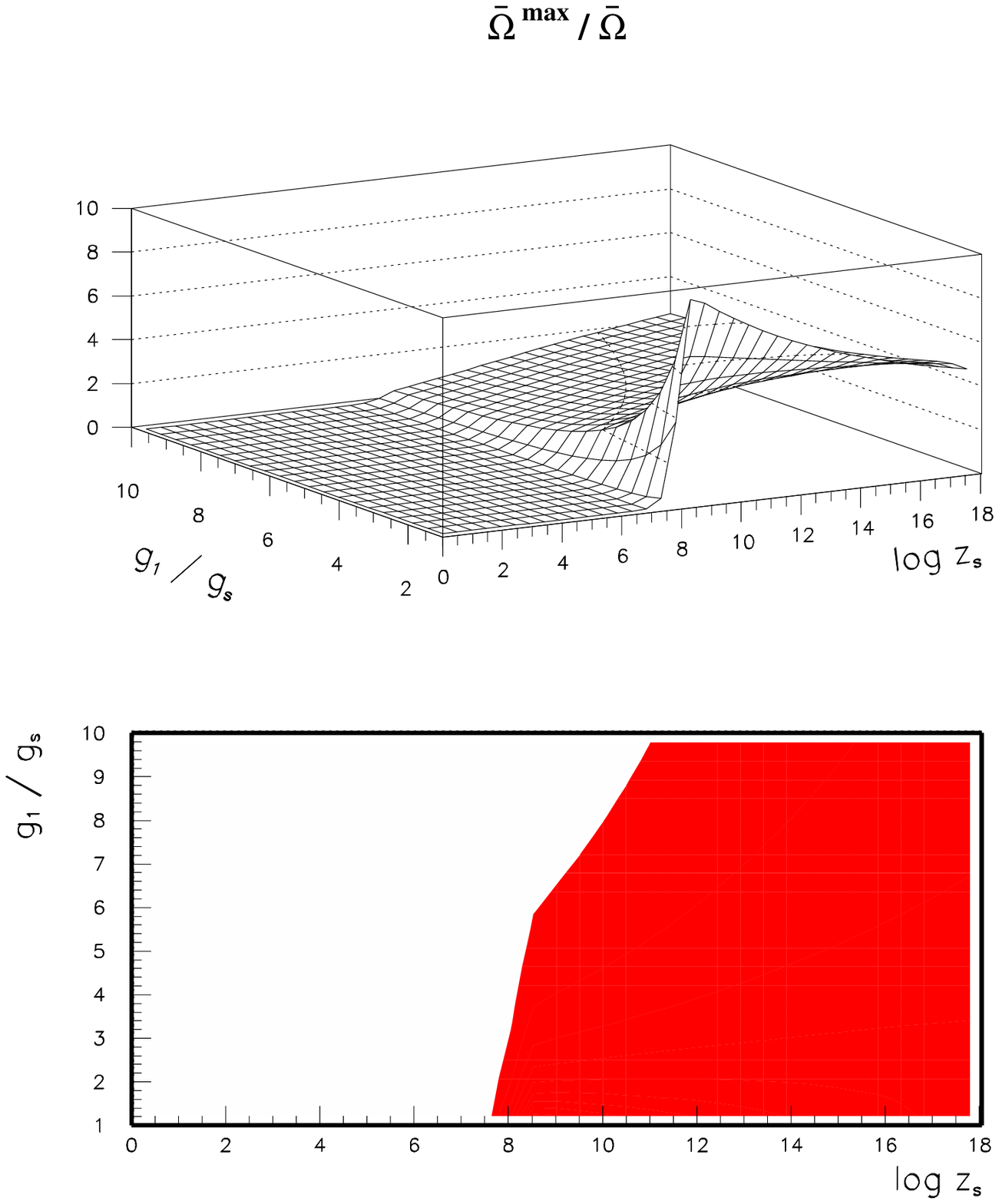}} &
      \hbox{\epsfxsize = 6.2 cm  \epsffile{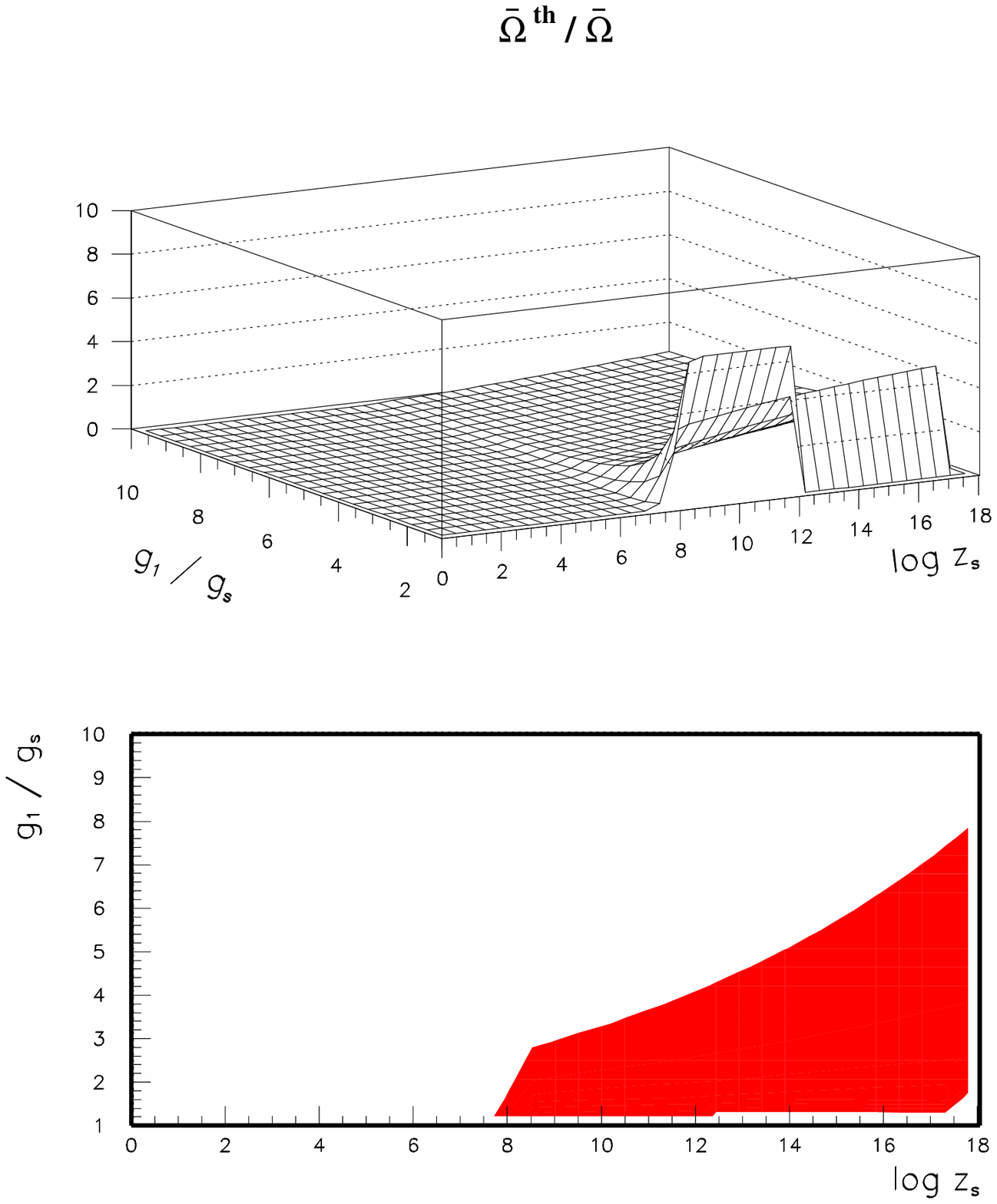}} \\
\end{tabular}
\end{center}
\vspace*{-1.5cm}
\caption[a]{We report the result of selective reduction in the case
where the  noise cause by the pendulum's internal modes is reduced
by a factor of ten, whereas the  pendulum and shot contributions
are left unchanged, i.e., $\vec{\rho}\,=\,(1, 0.1, 1)$.}
\label{noired2}
\end{figure}

Finally, for sake of completeness, we want to discuss the case of the
shot noise, i.e., the noise characteristic of the detector above 500
Hz. If
the shot noise is reduced by one tenth (i.e. $\rho_3\,=\,0.1$)  the
visibility region does not increase significantly. This result
is consequence of the fact that, as shown by Fig. \ref{noise}, the
shot noise contribution to the NPS starts to be relevant for
$f\,\sim\,1$ kHz, i.e., in a frequency region where the overlap between
the detectors begins to deteriorate (see Fig. \ref{over}).
In Figs. \ref{noired1} and \ref{noired2} the thermal noise is reduced
by one tenth and the increase in the visibility region is, comparatively,
larger. This shows that a reduction in the shot noise
will lead to an effect whose practical relevance is already questionable
at the level of our analysis.

Clearly,
the simultaneous reduction of  both  components of the thermal
noise leads to a substantial increase in the area of the visibility
region which gets even larger than
the ones illustrated in Fig. \ref{noired1}
and  \ref{noired2}.

\renewcommand{\theequation}{5.\arabic{equation}}
\setcounter{equation}{0}
\section{Discussion and executive summary}

There are no compelling reasons why one should not consider
the appealing theoretical possibility of a second VIRGO detector
coaligned with the first one. Moreover, recent experimental
suggestions seem coherently directed towards this goal \cite{gia}.
While the location of the second detector is still under
debate we presented a theoretical analysis of some of the
scientific opportunities suggested by this proposal.

We focused our attention on possible cosmological sources
of relic gravitons and we limited our attention to the case
of stochastic and isotropic background produced by the adiabatic
variation of the backgound geometry. In the framework of these
models we can certainly argue that in order to have a large
signal in the frequency window covered by VIRGO we have to focus
our attention on models where the logarithmic energy spectrum
increases at large frequencies. Alternatively we have to look
for models where the logarithmic energy spectrum exhibits some
bump in the vicinity of the VIRGO  operating window.
If the logarithmic energy spectra are decreasing as a function
of the present frequency (as it happens in ordinary inflationary
models) the large scale (CMB) constraints forbid a large signal
at high frequencies. In the case of string cosmological models
the situation seems more rosy and, therefore, we use these models
as a theoretical laboratory in order to investigate, in a
specific model the possible improvements of a possible VIRGO pair.
The choice of a specific model is, in some sense, mandatory. In
fact, owing to the form of the SNR we can immediately see that
different models lead to different SNR not only because the amplitude
of the signal differs in different models. Indeed, one can convince
himself that two models with the same amplitude at $100$ Hz but different
spectral behaviors between 2 Hz and 10 kHz lead to different SNR.

In order to analyze the sensitivity of the VIRGO pair we  described
a semi-analytical technique whose main advantage is to produce the
sensitivity of the VIRGO pair to a theoretical spectrum of arbitrary
slopes and amplitudes. The theoretical error is estimated, in our
approach, by requiring the compatibility with all the phenomenological
bounds applicable to the graviton spectra. As an intersting example,
we asked what is the sensitivity of a VIRGO pair to string cosmological
spectra {\em assuming} that a second VIRGO detector (coaligned with
the first one) is built in a european site. By assuming that the second
VIRGO detector has the same features of the first one we computed the
SNR and the related sensitivity achievable after one year of observation
in the case of string cosmological spectra.

By using the string cosmological spectra as a theoretical laboratory
we then studied some possible noise reduction. Our main goal, in this
respect, has been to spot what kind of stationary and stochastic noise
should be reduced in order to increase the visibility region of the
VIRGO pair in the parameter space of the theoretical models under
considerations. Our main result is that a selective reduction of each
of the three main sources of noise is not equivalent. A reduction in
the shot noise by a factor of ten does not increase significantly the
visibility region of the VIRGO pair. A selective reduction of
the thermal noise components is far more efficient. In particular,
we could see that a reduction (of one tenth) of the
pendulum's internal modes increases the visibility region of four
times. The simultaneous reduction of the  two components of the
thermal noise leads to an even more relevant increase.

The construction of a second VIRGO detector coaligned with the first
one and an overall reduction of the thermal noise of each detector
of the pair leads to what we called ``upgraded VIRGO'' program. The
results presented in this paper are obtained in the case of a
particularly promising class of theoretical models but can be generally
applied to any logarithmic energy spectrum with similar qualitative
results. However, owing to the non-linearities present in the evaluation
of the SNR it would not be correct assess that they hold, quantitatively,
without change. We hope that our results and our suggestions may turn out
to be useful in the actual process of design of the upgraded VIRGO program
\cite{gia}.

\section{Acknowledgements}
We would like to thank A. Giazotto for very
useful hints and for his kind interest in this investigation.

\end{document}